\documentclass[pmlr]{jmlr}

\makeatletter
\def\set@curr@file#1{\def\@curr@file{#1}} 
\makeatother
\usepackage[load-configurations=version-1]{siunitx} 

\usepackage{lipsum}
\usepackage{mwe}

\usepackage{tikz}

\makeatletter
\newcommand*{\centernot}{%
  \mathpalette\@centernot
}
\def\@centernot#1#2{%
  \mathrel{%
    \rlap{%
      \settowidth\dimen@{$\m@th#1{#2}$}%
      \kern.5\dimen@
      \settowidth\dimen@{$\m@th#1=$}%
      \kern-.5\dimen@
      $\m@th#1\not$%
    }%
    {#2}%
  }%
}
\makeatother

\newcommand{\indep}{\perp\mkern-9.5mu\perp}
\newcommand{\noindep}{\centernot{\indep}}

\usepackage{mathtools}
\newcommand\givenbase[1][]{\:#1\lvert\:}
\let\given\givenbase
\newcommand\sgiven{\givenbase[\delimsize]}
\DeclarePairedDelimiterX\Basics[1](){\let\given\sgiven #1}

\usepackage{algorithm}
\usepackage{algorithmic}
\usepackage{enumitem}
\usepackage{multirow}

\usepackage{longtable}

\usepackage{booktabs}
\usepackage[load-configurations=version-1]{siunitx} 

\theorembodyfont{\upshape}
\theoremheaderfont{\scshape}
\theorempostheader{:}
\theoremsep{\newline}

\jmlrvolume{106}
\jmlryear{2019}
\jmlrworkshop{Machine Learning for Healthcare}

\title[Causally Formulated Hazard Ratio Estimation]{A Causally Formulated Hazard Ratio Estimation through Backdoor Adjustment on Structural Causal Model}

\author{\Name{Riddhiman Adib} \Email{riddhiman.adib@marquette.edu} \\
      \addr Department of Computer Science\\
      Marquette University\\
      Milwaukee, Wisconsin, USA
      \AND
      \Name{Paul Griffin} \Email{paulgriffin@purdue.edu} \\
      \addr Regenstrief Center for Healthcare Engineering\\
      Purdue University\\
      West Lafayette, Indiana, USA
      \AND
      \Name{Sheikh Iqbal Ahamed} \Email{sheikh.ahamed@marquette.edu} \\
      \addr Department of Computer Science\\
      Marquette University\\
      Milwaukee, Wisconsin, USA
      \AND
      \Name{Mohammad Adibuzzaman} \Email{madibuzz@purdue.edu} \\
      \addr Regenstrief Center for Healthcare Engineering\\
      Purdue University\\
      West Lafayette, Indiana, USA
}

\begin{document}

\maketitle


\begin{abstract}
    Identifying causal relationships for a treatment intervention is a fundamental problem in health sciences. Randomized controlled trials (RCTs) are considered the gold standard for identifying causal relationships. However, recent advancements in the theory of causal inference based on the foundations of structural causal models (SCMs) have allowed the identification of causal relationships from observational data, under certain assumptions. Survival analysis provides standard measures, such as the hazard ratio, to quantify the effects of an intervention. While hazard ratios are widely used in clinical and epidemiological studies for RCTs, a principled approach does not exist to compute hazard ratios for observational studies with SCMs. In this work, we review existing approaches to compute hazard ratios as well as their causal interpretation, if it exists. We also propose a novel approach to compute hazard ratios from observational studies using backdoor adjustment through SCMs and do-calculus. Finally, we evaluate the approach using experimental data for Ewing's sarcoma.
\end{abstract}
\section{Introduction}
Experimental studies such as randomized controlled trials (RCT) are considered the gold-standard in hypothesis testing. For safety and efficacy reasons and regulatory purposes, most new drugs or treatments are studied through RCTs \citep{greene2012reform}. RCTs provide the best mechanism to identify the causal effect of treatments or interventions, by adjusting for observed and unobserved confounders under the rubric of a potential outcome framework \citep{fisher1960design}. Despite clear advantages of RCTs in drug-trials, in practice, they are expensive, time-consuming, and not feasible in many cases due to ethical reasons. Other issues with RCTs include low recruitment rate, loss to follow-up, insufficient sample size, and being prone to selection bias \citep{nichol2010challenging, frieden2017evidence}. While RCTs remain the best way to establish causation, large amounts of data captured with new technologies during routine healthcare (e.g., electronic health records (EHR) or wearable devices), colloquially termed \textit{big health data}, has the potential to discover causal effects from observational studies to complement RCTs. With proper methodological considerations, observational studies can provide a way to \textit{emulate} RCTs and go beyond statistical correlation \citep{hernan2008observational,hernan2017observational}. 

In the 1970s, the potential outcome framework was extended to observational studies to identify causal relationships from observational data through the Rubin Causal Model (RCM) \citep{rubin1974estimating, rosenbaum1983central,holland1986statistics}. Recent advances in structural causal model (SCM) provides the methodological framework under the potential outcome framework for graphically formalizing the identification of causal effects from observational and experimental data \citep{pearl2009causality,pearl2016causal}. SCMs can be used to \textit{emulate} RCTs from observational data in many cases if the graphical model is identifiable \citep{bareinboim2016causal}, which signifies the capability of estimating the interventional distribution ($P(y|do(x))$) from the available data with the assumptions incorporated in the model.  

Experimental studies (including RCTs) frequently explore and report survival analysis measures. Survival analysis is the branch of statistics that analyzes the expected duration of time-to-event with outcome statistics such as hazard ratio, odds ratio, and risk ratio. Survival analysis has been well-studied under the potential outcome framework with experimental studies and with RCM for observational studies \citep{cole2004adjusted,hernan2010hazards}. Recent research has also studied survival analysis with RCM for observational studies considering the data generating mechanism or the study designs to estimate outcome statistics such as hazard ratio, odds ratio, risk ratio, and risk difference \citep{didelez2007mendelian,hernan2004definition}. 

Commonly reported outcomes from survival analysis in experimental clinical studies include the survival curve and hazard ratio (HR). The survival curve graphically reports the hazard in a population and represents the fraction of the population that survived in the treatment and the control group over time. HR describes the comparative hazard between the treatment and the control group. Hazard function, or simply \textit{hazard} signifies the rate of events-of-interest (e.g., a death) at time $t$, conditional on survival until time $t$ and beyond \citep{spruance2004hazard}. Even though HR is widely used in practice as a standard tool for comparative evaluation of the outcome between treatment and control groups, it depends on the length of the study and, by definition, has an inherent selection bias (since only the survived population at time $t$ are \textit{selected} at time $t+1$) \citep{hernan2010hazards}. In addition, both the survival curve and HR do not consider the study design, that is RCT versus observational study, in their formalization. Consequently, it is difficult to interpret the results of an intervention from only the reported hazard ratio \citep{hernan2010hazards} and compare different studies with varying study designs and time lengths. The researcher has to consider the design of the study, length of the study as well has the hazard ratio to understand the effectiveness of the treatment. Structural causal models (SCMs) provide a framework to explicitly define the design of the study, the assumptions for the study, as well as the length of the study. However, to the best of our knowledge, a framework to compute the hazard ratio with SCMs does not exist. 

Previous approaches for adjusted survival curves under the rubric of RCM used inverse probability weighting (IPW) to adjust for confounders in the estimand \citep{cole2004adjusted}. However, this approach has a strong assumption, namely \textit{ignorability}  \citep{rubin1978bayesian,angrist1996identification}. The ignorability assumption states that there are no unobserved confounders in the model, and the variables considered for IPW satisfy the backdoor criterion. Although an approach with instrumental variable can be used when the treatment assignment is non-ignorable \citep{angrist1996identification}, in practice, this is rather a strong assumption and a variable can be a mediator, a collider, an M-bias,  or a confounder \citep{lederer2019control}. In this paper, we formulate the estimation of the hazard ratio from observational studies under the rubric of SCMs that does not depend on the ignorability assumption. We provide a principled approach to define observational studies using SCMs, redefine with time-specific survival as outcomes (instead of survival time as the only outcome), and therefore mathematically transform observational studies to the corresponding experimental studies by adjusting for confounders with the backdoor criterion and then, sample from the experimental studies to estimate hazard ratios. We provide the mathematical formalization of the approach with a simple causal graph and with detailed mathematical derivation, and validate the results with a simulated data set and a benchmark data set on Ewing's sarcoma.

\subsection{Clinical Relevance}
Most clinical research reports HR with survival analysis. However, the reported HR and its process of calculation do not take into account the study design (e.g., RCT vs. observational study) and corresponding assumptions (e.g., ignorability). This makes it harder to compare the results of different studies with different study designs, sample populations, study lengths and assumptions. Our proposed method with SCMs estimates HR by explicitly describing the study designs and assumptions for a better clinical understanding of the effect of the treatment of interest. 

\subsection{Technical Significance}
We propose a novel approach to estimate the HR from observational studies with SCM, taking the causal relationship between treatment and outcome into account. In HR calculation for survival analysis of observational studies, our review of the literature identifies a lack of causal interpretation. Our proposed approach first develops a time-invariant causal model and estimates the survival time after adjusting for the confounders in the SCM using backdoor adjustment and do-calculus. The development of an SCM enables us to identify the confounding variables, unlike with the ignorability assumption where we adjust for every variable available (except treatment and outcome), as well as properly adjust using the minimal set, thus reducing computational requirements. The computed survival times are considered ``as-if'' they were sampled from an RCT. The newly adjusted survival times are capable of expressing the true causal effect of treatment on the outcome through the survival curve and HR. We validate the proposed method in both simulated experiments and with observational data.

\subsection{Generalizable Insights}
We propose a novel method of estimating the HR for observational studies under the rubric of SCMs. The method can be used for any observational studies with survival data, after defining the SCM. Our method of estimating the HR through SCMs clearly defines the study-design and assumptions in the model. All the source code for this study is shared with the research community through a GIT repository. A Python-based library has been released that takes the data, the graph, and length of the study as input and provides the adjusted survival curve with backdoor adjustment and the hazard ratio as the output. Our approach is limited in the cases when i) the SCM is not defined and ii) the SCM is not identifiable through the adjustment formula or backdoor adjustment (i.e., there is no backdoor set).

\section{Related Work}
Survival analysis \citep{kleinbaum2010survival} is a methodological approach for modeling and comparing the time-to-event between two populations. The event is called a hazard, which can be death, an adverse clinical event, or a mechanical failure for physical systems. It compares the condition of survival in the treatment versus control group, and reports outcomes with statistical measures such as the HR. Frequently reported approaches in survival analysis include Kaplan Meier survival curve, Cox proportional hazards model, life tables, and survival trees,. We review a non-parametric approach of the Kaplan Meier survival curve and the semi-parametric approach of the Cox proportional hazards model.

The Kaplan Meier survival curve \citep{kaplan1958nonparametric} is a non-parametric statistic representing the survival function and HR in the treatment and the control group. It provides a visual comparison between survival functions in different treatment or control groups; it does not differentiate between RCTs or observational studies. Data from both of the approaches can be plotted as the Kaplan Meier curve. It is up to the individual researcher to interpret and explain the Kaplan Meier curve based on the study design. Cox PH model, on the other hand, is computationally complex. However, it is a commonly used approach for survival analysis, and is widely used to compute the HR in epidemiological studies. The key aspect of it is the underlying proportional hazards assumption \citep{cox1972regression}, stating that the HRs of the treatment and control group are proportional and is a function of the covariates. It is a semi-parametric model since no assumption is made about the baseline hazard function (i.e., hazard function with no covariates). In general, it is effective in estimating both regression coefficients ($\beta_i$) and the HR \citep{kleinbaum2010survival}.  Futher, it is unbiased \textit{(when estimated considering all possible covariates)}. 

We review existing approaches to compute the HR for observational and experimental studies. Previous work on survival analysis for observational data with RCM under potential outcome used IPW to adjust for confounders \citep{cole2004adjusted}. However, RCM requires the ignorability assumption that all variables considered for adjustment with IPW satisfy the backdoor criterion. In reality, a variable can also be a mediator, and in those cases adjusting for the mediators will result in inaccurate analysis. It has also been shown that the HR estimation approach has an inherent selection bias \citep{hernan2004structural,hernan2010hazards} as only the patients who survived at time $t$ were sampled at time $t+1$ to be considered for the estimation. SCMs provide the mathematical machinery to identify the backdoor variables given a causal graph. We used the same Ewing's sarcoma data set as studied in \citep{cole2004adjusted} with the same assumptions (i.e., all the covariates satisfy the backdoor criterion) to arrive at the same result as a validation strategy for our approach. 

For survival analysis, it was shown that in some cases the Kaplan Meier curve may show no difference between the  treatment and control groups when in reality there is a statistically significant difference in the HR, if it is  adjusted properly \citep{makuch1982adjusted}. The rationale behind this phenomenon is that a non-parametric approach is used to plot the survival curve, whereas a semi-parametric method is used to calculate the HR. The authors \citep{makuch1982adjusted} presented an approach to construct a plot of the survival curve consistent with the HR calculated. In this work, the adjusted survival curve for a specific treatment group was estimated by calculating a mixture of the estimated survival functions for separate strata, and weighted based on the distribution of the covariate in the sample dataset.  However, the approach does not consider the design of the study in the survival analysis. 

To extend the existing definition of the Cox PH Model, the Marginal Structural Cox PH Model has been introduced and used to find the effect of Zidovudine on the survival of HIV-positive men \citep{hernan2000marginal}. Statistical analysis in the presence of time-dependent confounders is commonly done through a standard Cox PH model.  However, Robin \citep{robins1997causal} has previously shown that this approach cannot adjust for all biases. Similar to previous work under the RCM, the authors used the conditional ignorability assumption. This is a much stronger assumption compared to using the SCM to identify confounding variables opening the backdoor. Several other researchers  \citep{satten2001kaplan,rotnitzky2014inverse} have used the IPW approach, although without using SCMs.

The existing literature to compute the HR does not consider the study design and might lead to misinterpretation if the data were not sampled correctly or adjusted for the right confounding variables. While previous research alludes to this problem, they do not provide the mathematical machinery for survival analysis. Although the traditional Cox PH Model can minimize the effects of biases, it is not the same as ``adjustment" of confounding variables. The bias is reduced by fitting the Cox PH regression model until convergence \citep{cox1972regression}, it does not consider the study design or the data generating mechanism. The model fitting approach does not generate a causally meaningful interpretation despite reduction in biases. Our goal is to formulate an approach that estimates the HR through a causal formulation considering the data generating mechanism with SCM, that portrays the direct causal effect of treatment on outcome, in terms of the HR metric. The assumption of variables opening the backdoor path in the SCM as confounders and adjustment on the dataset based on that enables a more causally interpretable estimation of the HR. 
\section{Background}
\subsection{Hazard Ratio}

To define the HR, we use the hazard function \citep{spruance2004hazard} in the Cox proportional hazard model:

\begin{equation}
    h(t, \mathbf{X}) = h_0(t) \exp(\sum_{i=1}^p \beta_i X_i)\label{eq:h(t,X)}
\end{equation}

Based on this, the Hazard Ratio (HR) is defined  \citep{kleinbaum2010survival} as: 

\begin{equation}
    HR = \frac{h(t, \mathbf{X}_{x=1})}{h(t, \mathbf{X}_{x=0})}\label{eq:HR}
\end{equation}

Here, $h(t, \mathbf{X})$ represents the hazard function at time $t$ and the vector with the covariates of the model $\mathbf{X}$. $\mathbf{X}$ can also be written as $[w_0, w_1, ..., w_m, z_0, z_1, ..., z_n, x]$, where $x$ is the treatment, $z_i$ are the confounders, and $w_i$ are the other associated covariates. $\mathbf{X}_{x=1}$ represents the value of the covariate vector $\mathbf{X}$ with value of the treatment set as 1 ($x=1$), making $\mathbf{X}_{x=1} = [w_0, w_1, ..., w_m, z_0, z_1, ..., z_n, 1]$. $\beta$ represents the \textit{maximum likelihood estimates (MLE)} for each covariate. In other words, $\beta$ is the corresponding coefficient for each covariate that fits the data into a converging model for the Cox regression. 

As expressed in \autoref{eq:h(t,X)}, the proportional hazard assumption defines the hazard function $h(t, \mathbf{X})$ to be composed of the baseline hazard function $h_0(t)$ (i.e., hazard when all covariates are set to 0), multiplied with the exponential of the sum of $\beta$ multiplied by the corresponding covariate.

Since we have defined the HR and hazard function, we can simplify the equation of the HR to: 

\begin{equation}
\begin{aligned}
    HR & = \frac{h(t, \mathbf{X}_{x=1})}{h(t, \mathbf{X}_{x=0})} \\
      & = \frac{h_0(t) \exp(\beta_x 1 + \beta_z Z + ...)}{h_0(t) \exp(\beta_x 0 + \beta_z Z + ...)} \\
      & = \exp(\beta_x)
\end{aligned}
\end{equation}
In other words, the HR is equivalent to the exponential of the regression coefficient $\beta$. However, computing $\beta$ is non-trivial since, in practice, one does not know the baseline hazard function \textit{($h_0(t)$)}. We can only estimate the HR using the maximum likelihood function, and iterating until the model converges to a pre-defined error bound \citep{kleinbaum2010survival}.


Although the HR is an important outcome, it has limitations in explaining causal relationships. No causal mechanism is understood from the HR. This is because the HR is calculated from the convergence of regression models and, confounding and other such bias is handled by simply including the covariates to the model. It is then up to the individual researcher to make sure that the right data are used to measure the HR and interpret accordingly. For example, an HR calculated from an RCT provides the casually linked hazard for the intervention, whereas the same HR calculated from an observational study simply provides a correlated hazard. This existing approach simplifies the calculation and reduces the burden on the researcher. However, we frequently find differences between the survival curve and the HR. This difference, or bias, arises because of the inherent definitions of the survival curve and Cox PH model.

\subsection{Structural Causal Models}
Structural causal models (SCMs), developed on the foundations of probabilistic graphical models, draw inferences that explain the causal relationship between variables. With an SCM, a causal model is defined first and is expressed with a graphical representation. Definition 1 gives the formal description of an SCM: \citep{bareinboim2016causal,pearl2009causality}.

\begin{definition}[Structural Causal Model]
    A structural causal model $M$ is a 4-tuple \\$\langle U, V, f, P(u) \rangle $ where:
    \begin{enumerate}
        \item $U$ is a set of background (exogenous) variables that are determined by factors outside of the model,
        \item V is a set $\{V_1, V_2, ..., V_n\}$ of observable (endogenous) variables that are determined by variables in the model (i.e., determined by variables in $U \cup V$ ),
        \item F is a set of functions $\{f_1, f_2, ..., f_n\}$ such that each $f_i$ is a mapping from the respective domains of $U_i \cup P A_i$ to $V_i$, where $U_i \subseteq U$ and $P A_i \subseteq V \ V_i$ and the entire set $F$ forms a mapping from $U$ to $V$. In other words, each $f_i$ in $v_i \leftarrow f_i(pa_i, u_i), i = 1, ..., n$, assigns a value to $V_i$ that depends on the values of the select set of variables $(U_i \cup P A_i)$, and
        \item $P(u)$ is a probability distribution over the exogenous variables.
    \end{enumerate}
\end{definition}

An SCM is often expressed by a causal graph $G$. Each node $V$ in $G$ represents an observed or unobserved variable, and each directed edge $E$ represents the causal relationships between them. To find the causal effect of variable $X$ on variable $Y$, do-calculus is introduced \citep{pearl2016causal}. Do-calculus is used to map the observational reality to the corresponding experimental reality with the identifiability equation by adjusting for different kinds of biases (e.g., confounding bias), if it exists. The backdoor criterion provides a powerful tool to identify the variables that need to be adjusted for this transformation (in other words, adjust for confounding bias) and is defined in definitions 2 and 3.

\begin{definition}[Backdoor Criterion]
    Given an ordered pair of variables $(X, Y)$ in a directed acyclic graph $G$, a set of variables $Z$ satisfies the backdoor criterion relative to $(X, Y)$ if no node in $Z$ is a descendant of $X$, and $Z$ blocks every path between $X$ and $Y$ that contains an arrow into $X$.
\end{definition}

\begin{definition}[Backdoor Adjustment]
    If a set of variables $Z$ satisfies the back-door criterion relative to $(X, Y)$, then the causal effect of $X$ on $Y$ is identifiable and is given by the formula: $P(y|do(x)) = \sum_z P(y|x,z) P(z)$
\end{definition}

\subsection{Problem Definition}
Our research problem is to develop a method to compute the HR for observational studies by leveraging the SCM by explicitly declaring our assumptions and adjusting for the right confounders. The goal is to acknowledge the defined roles of variables in the SCM, and use a minimum set of confounders to adjust for backdoor, thus building a computationally-efficient and more accurate model for objective estimation and comparison. The algorithm will take three sets/inputs, (1) observational dataset $D$ consisting of treatment, outcome in survival-time and other covariates, (2) SCM supporting the causal mechanism and dataset, $G$, and, (3) length-of-trial $T$. At the completion of the algorithm, the output will be: (1) adjusted survival curve $S$ \textit{(non-parametric estimation)}, and (2) hazard ratio of treatment, $HR$ \textit{(semi-parametric estimation)} (\autoref{fig:approach_schema}). The assumption in our approach is that the observational data are available, and the SCM is fully specified. 

\section{Methods}
\begin{figure}[htbp]
    \centering
    \includegraphics[scale=0.48]{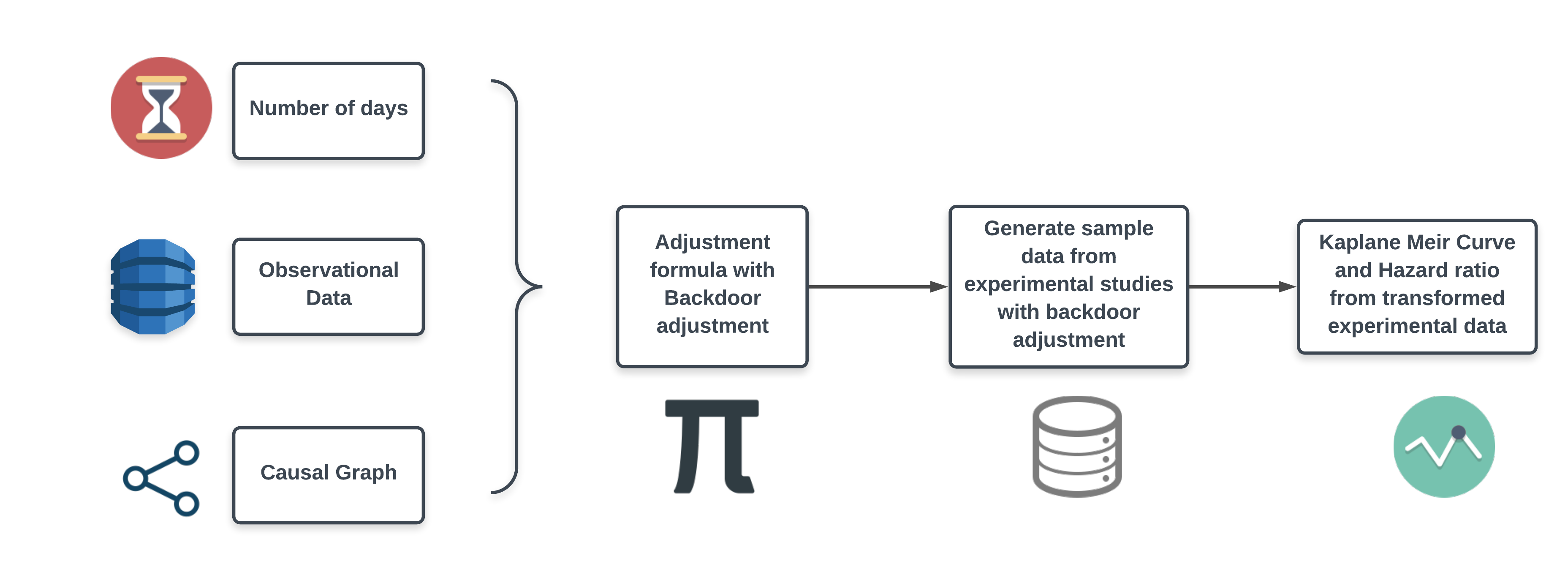}
    \caption{Schematic overview of the proposed approach. Observational data, corresponding causal diagram and the length of study is provided as input. The approach first use backdoor adjustment to create sample data from experimental study, and then compute the hazard ratio from the sampled experimental data.}
    \label{fig:approach_schema}
\end{figure}

In this section, we formalize our approach to mathematically transform the time-dependent observational data to the corresponding experimental data by leveraging the SCM. We then use the adjusted dataset for estimating HR using Cox PH Model. Our proposed approach focuses on causal effect of treatment on outcome to measure HR. We start with an observational study scenario and define all related assumptions. The schematic diagram for the proposed approach is shown in \autoref{fig:approach_schema}.

\subsection{Assumptions}
\begin{figure}[htbp]
    \centering
    \includegraphics[scale=0.25]{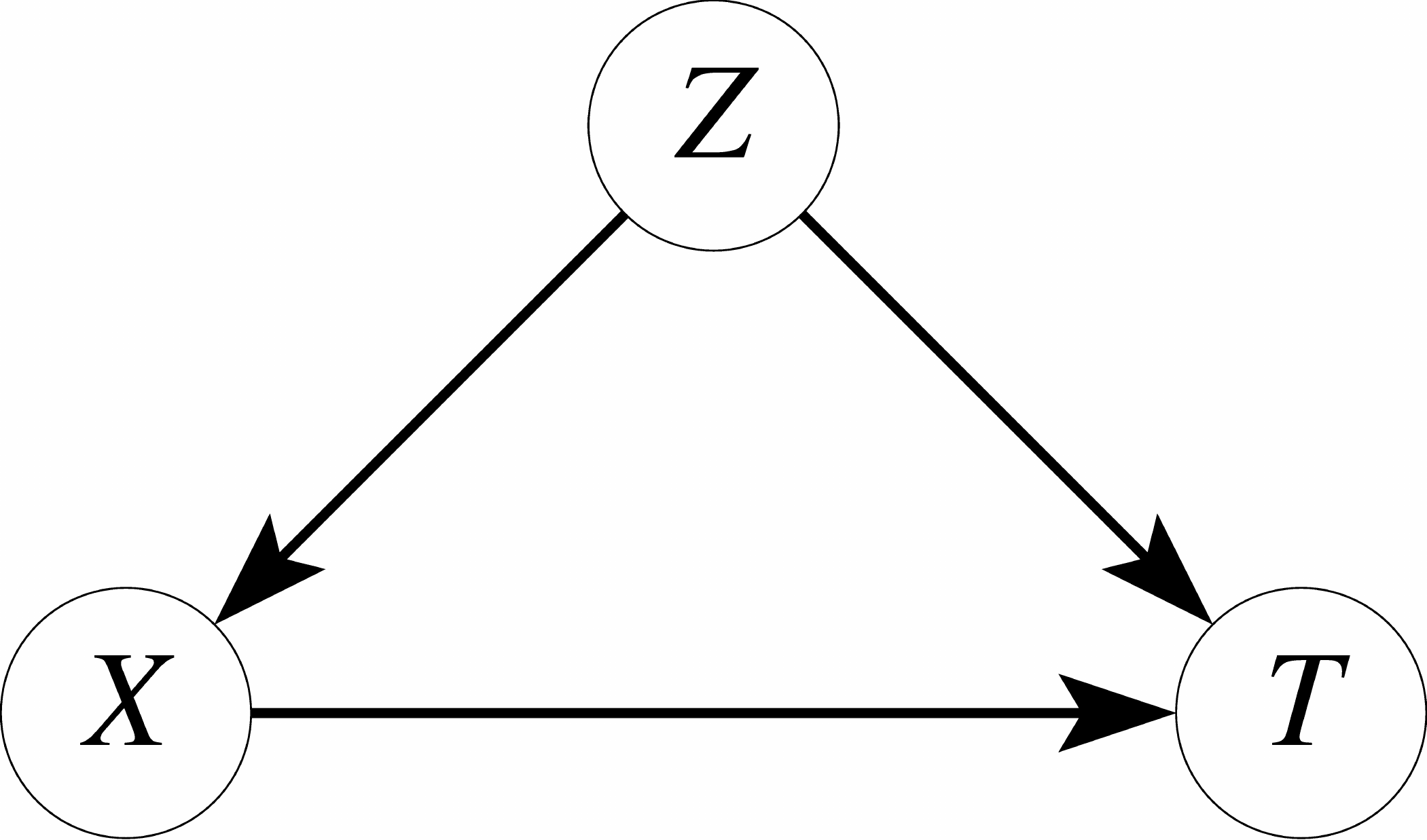}
    \caption{Simple observational study treatment $X$, outcome in survival-time $T$ and single confounder $Z$, expressed using causal directed acyclic graph \textit{(nodes are the variables, edges portray causal relationships between variables)}}
    \label{fig:simple_dag}
\end{figure}

We assume a simple observational study for a population, consisting of treatment $X$, confounding variable $Z$, and outcome in survival time $T$. In this scenario, treatment $X$ is a dichotomous variable ($X=1$ signifying treatment and $X=0$ signifying control). Outcome $T$ is the survival time from the beginning of the study and is a continuous variable in time units. Although the confounding variables can be a categorical or continuous variable, for simplicity, we assume the confounder $Z$ to be a dichotomous variable. This observational study can be expressed as an SCM and with a graphical form $G$ through causal directed acyclic graph (causal DAG) in \autoref{fig:simple_dag}, where treatment, confounder, and outcome is expressed by the nodes $X$, $Z$ and $T$ respectively.

From the definition of the SCM, we can express the underlying functions defining the causal relationships between variables by: $Z \leftarrow f_z(U_z)$, $X \leftarrow f_x(Z, U_x)$, $T \leftarrow f_t(Z, X, U_t, h_0(t))$. Here, $U = \{U_x, U_t, U_z\}$ is the set of exogenous variables, $V = \{Z, X, T\}$ is the set of endogenous variables, $f = \{f_z, f_x, f_t\}$ is the set of structural functions.

\begin{itemize}
    \item $f_z(U_z)$ shows that confounder $Z$ is independent of any other endogenous variables. 
    \item $f_x(Z, U_x)$ expresses the dependency of $X$ on $Z$. Since $Z$ is parameter for both functions $f_x$ and $f_y$, $Z$ imposes a bias on the model ($P(X|Z=0) \neq P(X|Z=1)$), and the function $f_x$ defines whether the bias is strong or weak.
    \item $f_t(Z, X, U_t, h_0(t))$ defines the effect of $X$ and $Z$ on the survival time $T$. This function also depends on the baseline hazard function $h_0(t, \mathbf{X})$ since this defines the rate of decline in survival.
\end{itemize}
We also assume to know the sample size of population $n$ and a maximum length of survival time $t_{max}$.

\subsection{Approach}

\subsubsection{Transformation of single study to multiple studies}

Experimental studies commonly have different study time-lengths, e.g., different number of days as the outcome endpoints (e.g., 30-day survival, 90-day survival, etc.). This variable is a dichotomous variable and describes a patients' status of survival at the end of the study. While analyzing a study similar to these, we do not take into account survival at each day, or survival after end-of-trial day, since we do not have the opportunity to do so. In our problem definition, we only have the survival time of individuals; however there is no defined end time for the trial. From the individual survival time, We can easily get the $i$-th day survival of every individual in the dataset, $i$ being the number of days from the beginning of the study. We use \textit{days} as smallest unit of time, since we assume the dataset reports survival in units of days. However, it could be any other units of times (\textit{e.g. minutes, or weeks}) depending on the problem domain and dataset.

Since our observational study has a maximum survival time of all individuals $t_{max}$, we assume we calculate the variables $Y_i$, signifying the $i$-th day survival, $i$ ranging from $0$ to $t_{max}$. Conversion of continuous variable $T$ describing survival time into multiple variables $Y_i$, each describing survival at the $i$-th day, essentially breaks down the single observational study into $t_{max}$ number of observational studies with variables $X$, $Z$ and $Y_i$, each of which is now a dichotomous variable.

\begin{figure}[htbp]
    \centering
    \begin{tabular}{ccc}
        For outcome $Y_0$ & For outcome $Y_1$ & For outcome $Y_n$ \\
        \includegraphics[scale=0.15]{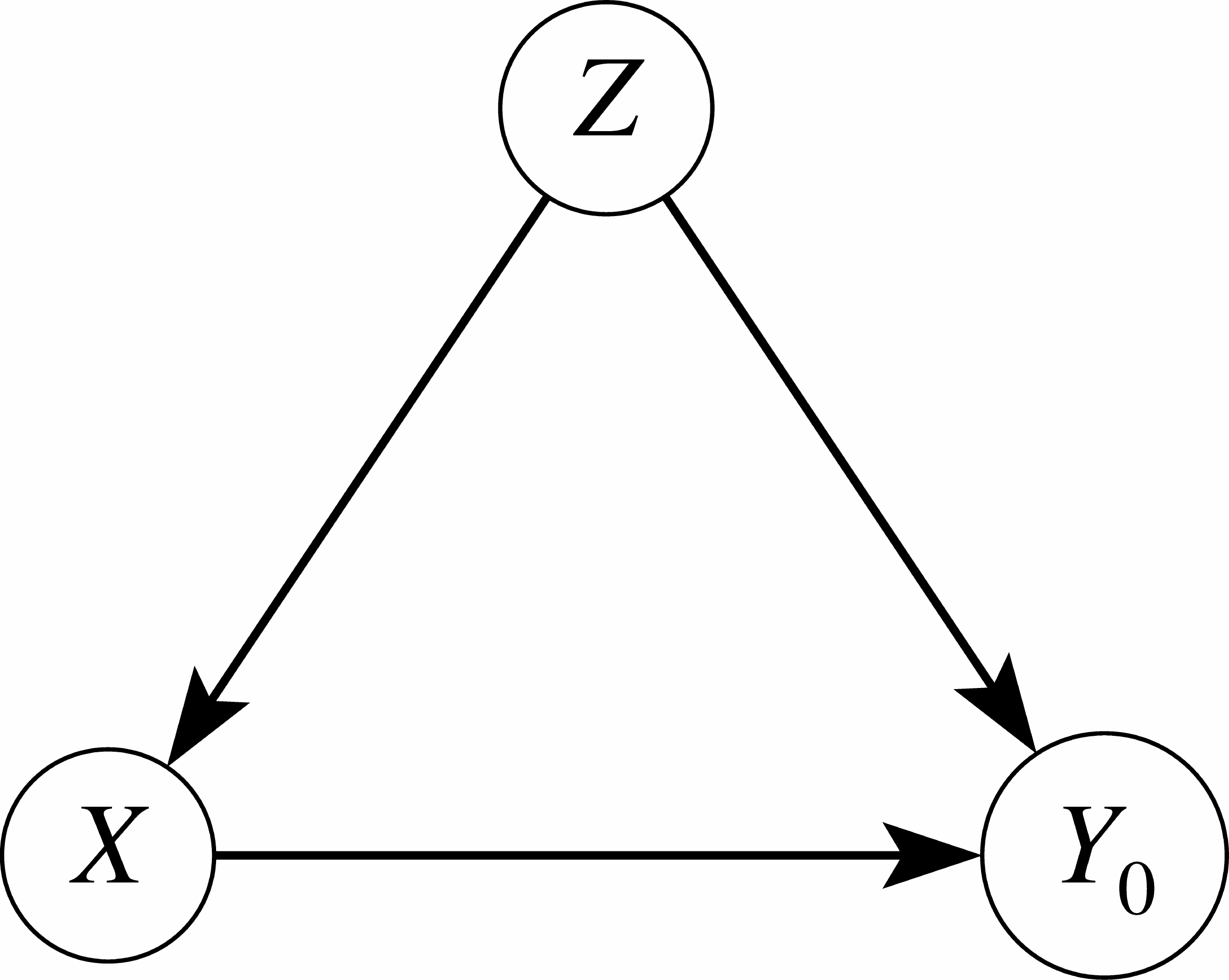} & \includegraphics[scale=0.15]{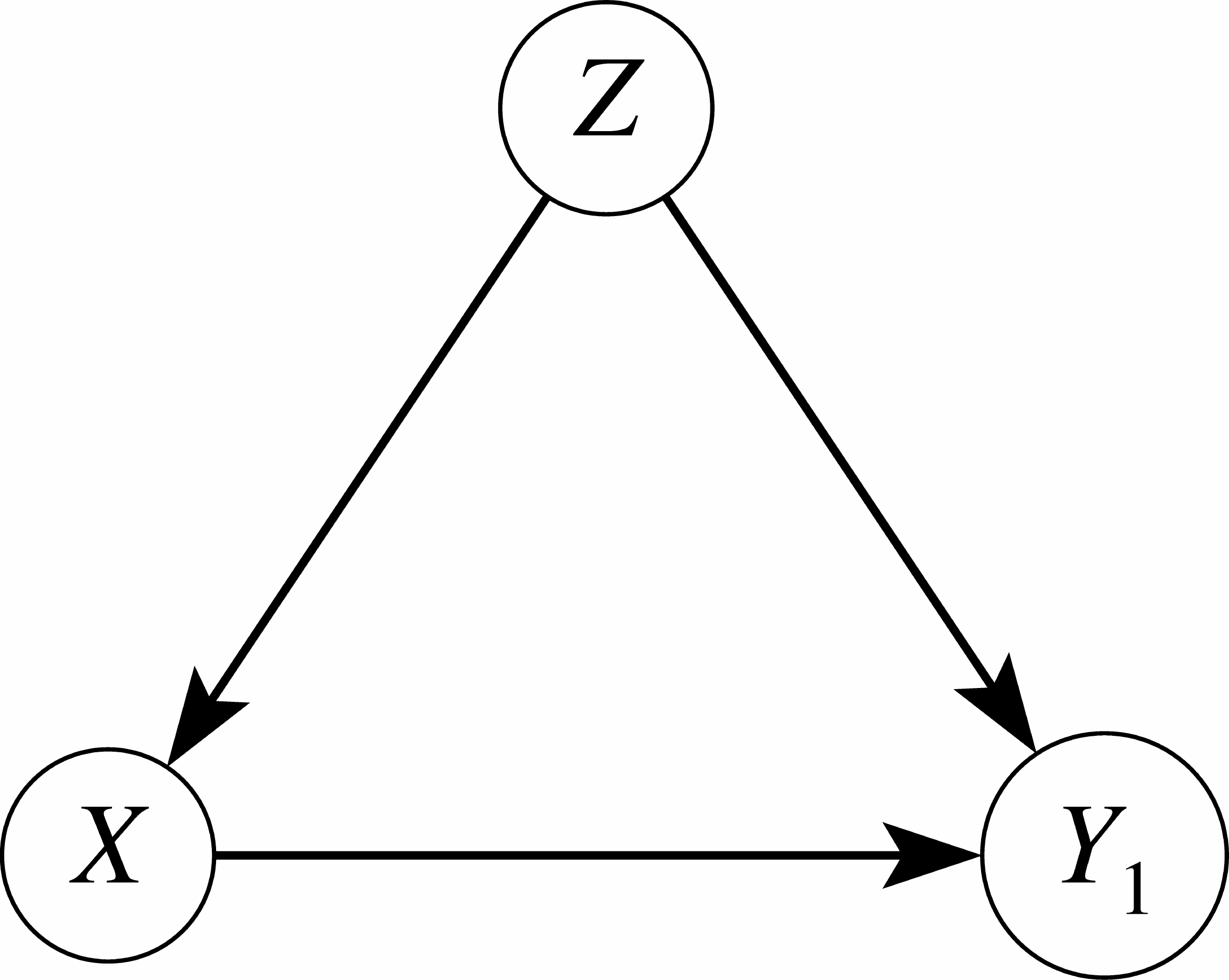} & \includegraphics[scale=0.15]{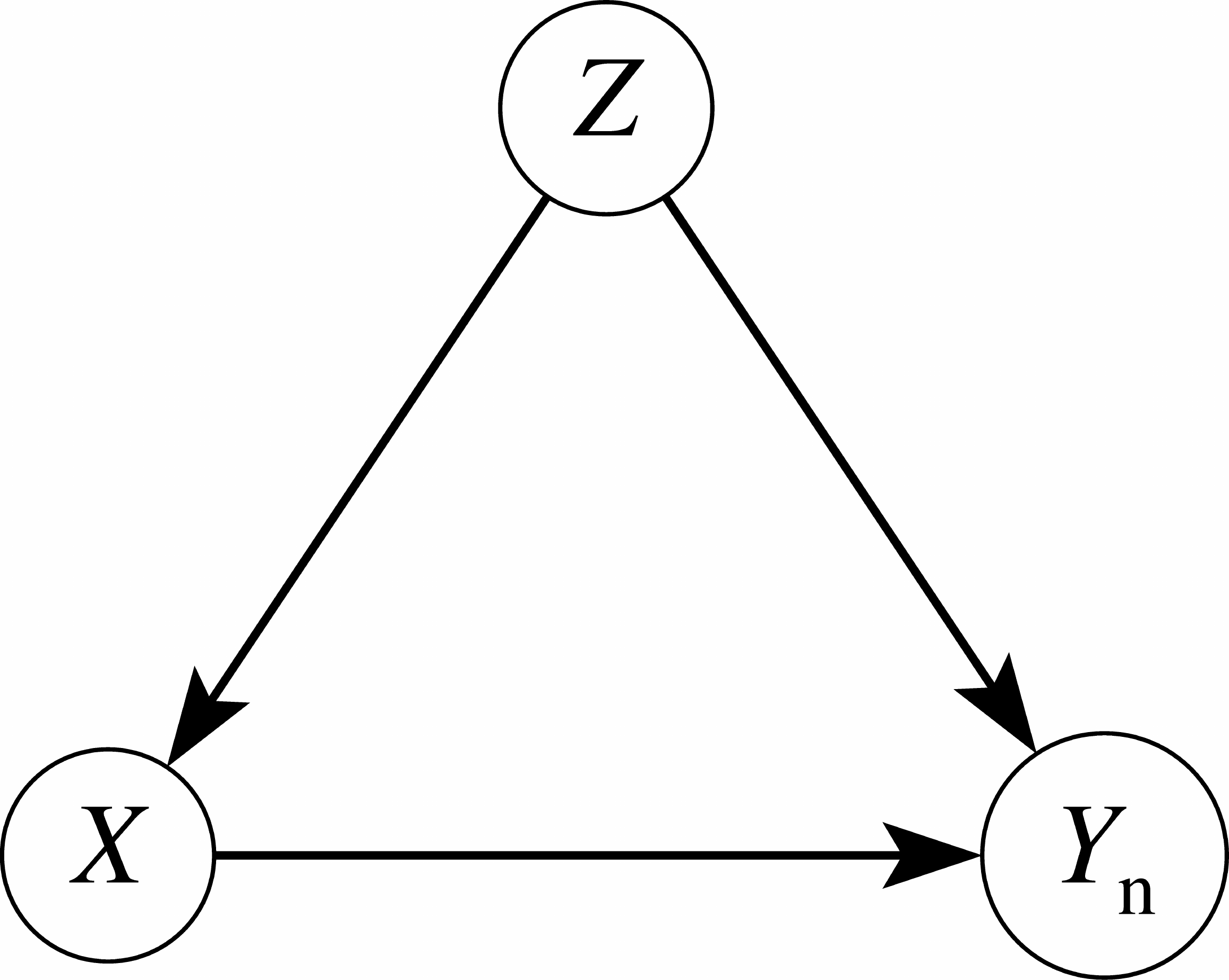}\\
        \multicolumn{3}{c}{(a) Converted Causal DAGs with no dependency between $Y_i$s} \\
        \multicolumn{3}{c}{\includegraphics[scale=0.4]{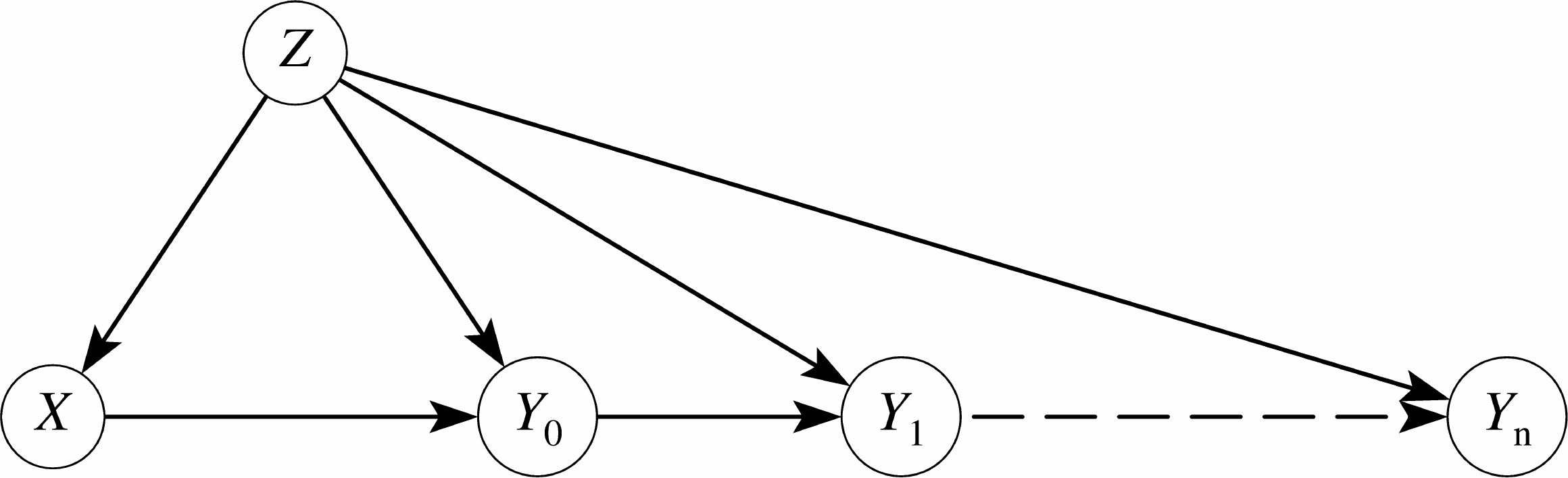}}\\
        \multicolumn{3}{c}{(b) Converted single Causal DAG with  dependencies between $Y_i$s} \\
    \end{tabular} 
    \caption{Converted Causal DAGs with survival time converted to binary outcome of survival at different timepoints}
    \label{fig:new_dags}
\end{figure}

Through the transformation, from a single SCM $G$, we end up with $n$ different SCMs, each with the same treatment $X$ and the confounder $Z$, but different outcome (survival at $i$-th day). Note that, in our assumption, the causal graph is time-invariant, i.e., the functional relationship between the variables does not change over time. This conversion is represented by $n$ different SCMs (Transformed graphs A, \autoref{fig:new_dags} (a)), where $n \geq t_{max}$.

An important point to note here is that, the single confounder $Z$ and treatment $X$ from the original observational study is not being transformed, only the outcome is distributed into multiple variables. In other words, we assume a point intervention and the confounding variables are invariant in time. And since we are transforming from a single trial to multiple trials, the outcomes $Y_i$s of these separate trials are not conditionally dependent on each other \textit{(e.g. a RCT with 30-day survival as outcome does not analyze about whether any patient died at 20th day or 29th day.)}. 

However, in extracting information from obs. data, there is dependency between them. Specifically, $Y_i$ has causal effect on $Y_{j}$ (where $j > i$), since if $Y_i$ is 0 \textit{(e.g. patient died at $i$-th day)}, all $Y_{j}$ (where $j > i$) is 0 \textit{(e.g. patient remains dead for all consecutive days)}. Also, $Y_i$ only has direct causal effect on $Y_{i+1}$, every other corresponding effect is mediated through. If $X$ has causal effect on $Y_i$, it is mediated through $Y_{i-1}$. For example, $X \indep Y_1 | Y_0$, in absence of any backdoor variables. The relationship between $Y_i$s is reflected through a single transformed SCM (Transformed graph B, \autoref{fig:new_dags} (b)), where $n \geq t_{max}$. The similarities between transformed graphs A and transformed graph B is that they both have same $Z$ and $X$, and the dissimilarities are:
\begin{enumerate}
    \item Transformed graphs A portrays $n$ different trials with different outcomes, whereas transformed graph B is a single trial.
    \item For transformed graphs A, $Y_i \indep Y_j$ (where $j \neq i$), however for transformed graph B, $Y_i \noindep Y_j$ (where $j \neq i$).
    \item Since two causal DAGs are different, transformed graphs A and transformed graph B have two different equations for $P(Y_i|do(X))$.
\end{enumerate}

In summary, we transform the single observational study into multiple different trials expressed through two different transformations (transformed graphs A and transformed graph B, \autoref{fig:new_dags}), each with the same treatment $X$ and confounding $Z$, but with different survival time as the outcomes, as the death (or failure) increases over time. These outcomes are the status of survival (or death) at $i$-th day, where $i$ is $0$ to $n$ ($n \geq t_{max}$). 



\subsubsection{Generation of Survival Curve}

Applying Backdoor adjustment formula in transformed graphs A, the causal effect of $X$ on $Y_i$ (for all $n$ causal graphs) is: $$P(Y_i|do(X)) = \sum_Z P(Y_i|X,Z) P(Z)$$
In transformed graph B, the causal effect of $X$ on $Y_i$ is: $$P(Y_i|do(X)) = \sum_{Z,Y_{i-1},...Y_1,Y_0} \left( \prod_{k=0}^n P(Y_k|Y_{k-1},...,Y_0,X,Z) \right) \cdot P(Z)$$
Since $P(A|B,C)P(B|C) = P(A,B|C)$ \textit{(using rules of conditional probabilities)}, we can reduce this equation to, $$P(Y_i|do(X)) = \sum_{Z,Y_{i-1},...Y_1,Y_0} P(Y_i,Y_{i-1},...,Y_0|X,Z) P(Z)$$
Finally, for $j<=i$, $P(Y_j=1|Y_i=1)=1$ \textit{(e.g. if a person is alive at 30th day, he has been alive for the last 29 days as well)}, $P(Y_i=1, Y_{i-1}=1) = P(Y_{i-1}=1|Y_i=1)P(Y_i=1) = P(Y_i=1)$, which reduces our equation down to the same as that of transformed graphs A: $$P(Y_i=1|do(X)) = \sum_{Z} P(Y_i=1|X,Z) P(Z)$$

This signifies whether we use transformed graphs A or transformed graph B, we end up with same adjustment formula.

For each of the newly transformed causal DAGs, we can now adjust for the confounder using the backdoor adjustment formula. We calculate adjusted probabilities $P_{adj}$ and thus adjusted counts $C_{adj}$ for each of the $n$ causal graphs. Using the values of $P_{adj}$, we generate survival curve with Kaplan Meier fitter. 

\subsubsection{Calculation of Hazard Ratio}
Since we calculated $C_{adj}$ for each of $n$ causal graphs, we know number of adjusted individuals alive at each unit (day) of time. This helps us build back the adjusted survival time $T_{adj}$ for individuals, as it was in the original dataset. The newly calculated survival time $T_{adj}$ is adjusted for the confounding bias, as if they were sampled from an RCT. We measure the HR using Cox PH model with the adjusted survival time $T_{adj}$ as outcome.

\subsubsection{Algorithm}



Algorithm 1 generates adjusted Kaplan Meier curve as well as the HR from Cox PH model on the adjusted dataset. The input for the algorithm is the dataset, specifically, confounder $Z$, treatment $X$, survival time $T$, and event status $S$. In the algorithm, variables in uppercase letters signify vectors, and variables in lowercase signify single variables. Internal procedures \textit{convert\_single\_to\_multiple\_trials} (Algorithm 2) are shown separately. 


\begin{algorithm}[tb]
\caption{Causally Formulated Hazard Ratio Estimation}
\label{alg:main}
\textbf{Input}: $Z$, $X$, $T$, $S$\\
\textbf{Output}: $survival\_curve$, ${HR}_{drug}$
    \begin{algorithmic}[1] 
        \STATE global $n \gets length(T)$
        \STATE global $t_{max} \gets max(T)$
        
        \STATE $Y_i \gets convert\_single\_to\_multiple\_trials(T, S)$
        
        \FOR {$i \gets 0$ to $t_{max}$}
            \STATE $adj\_p_i \gets \sum_{Z} P(Y_i=1\given X,Z) P(Z)$
            \STATE $adj\_c_i \gets adj\_p_i * count(X)$
        \ENDFOR
        
        \STATE $survival\_curve \gets plot(time, cumulative(adj\_p_i))$
        
        \STATE $adj\_X, adj\_T \gets convert\_multiple\_to\_single\_trial(adj\_c_i)$
        
        \STATE $ model \gets cox\_ph\_model(adj\_X, adj\_T)$
        \STATE $HR_{drug} \gets exp(model.\beta_{drug})$
        \STATE \textbf{return} $survival\_curve, HR_{drug}$
    \end{algorithmic}
\end{algorithm}

\begin{algorithm}[tb]
\caption{Conversion of single trial to multiple trials}
\label{alg:step1}
\textbf{Input}: $T$, $S$\\
\textbf{Output}: $Y_i$
    \begin{algorithmic}[1]
        \FOR{$i \gets 0$ to $t_{max}$}
            \FOR{$j \gets 0$ to $n$}
                \STATE $Y_i[j] \gets (T[j] <= i$) \AND ($S[j] = 1)$ ? 0 : 1
            \ENDFOR
        \ENDFOR
        \STATE \textbf{return} $Y_i$
    \end{algorithmic}
\end{algorithm}

\section{Experiments and Applications}
We evaluated the proposed approach for computing HR and visualizing survival curves with an experimental and observational dataset: (1) a synthetic dataset derived from a linear acyclic model with Gaussian noise; (2) a real-world dataset on disease-free survival in patients with Ewing's Sarcoma \citep{makuch1982adjusted}. The rationale to consider these two datasets are: (1) both of the underlying causal model has a backdoor path through confounders, and, (2) both these datasets have treatment and control group that satisfy the proportionality hazarads assumption.


\subsection{Experimental Data}
We simulate an observational study with $n=200$ patients. A subgroup of the patients received a treatment ($X=1$), and the remaining patients did not ($X=0$). We generate data on survival time $T$ (in days) defined as the outcome. The treatment assignment is confounded by sex (e.g. $Z=1$ for female, $Z=0$ for others). The scenario has a causal model as depicted in \autoref{fig:simple_dag}. 

For the simulation, we generated outcome variable survival-time through defining a baseline hazard function. We defined survival time to be exponentially varying with time, in the form of: $T \gets a . exp((b + c Z + d X + e Z X) * i) + E$, with $Z$ being confounder, $X$ being treatment, $E$ being the noise/error and $i$ being the index of patient. The other parameters were set to $a = 5, b = 0.025, c = 0.005, d = - 0.015, e = 0.075, E = U(-0.5, 0.5)$, they were selected such that the HR remains close to 1, however injection of bias through $Z$ portrays different outcome in survival curve.

We simulate the study with a strong biased effect from confounder $Z$. We define the strength of bias by an imbalance of conditional probabilities in each stratum of $Z$ through the function $f_x(Z, U_x)$. For the defined strong bias case, $P(X=1|Z=0)=0.75$ and $P(X=1|Z=1)=0.25$. It translates to, if $Z=0$ stands for females in this trial, 75\% received the drug, whereas, in $Z=1$ or others, only 25\% received the drug. In a randomized controlled trial, under a no-confounding-bias scenario, we should have $P(X=1|Z=0)=P(X=1|Z=1)=0.5$.

\begin{figure}[t] 
    \centering
    \begin{minipage}{0.45\linewidth}
        \centering
        \includegraphics[width=\linewidth,trim= 10 1 30 34, clip]{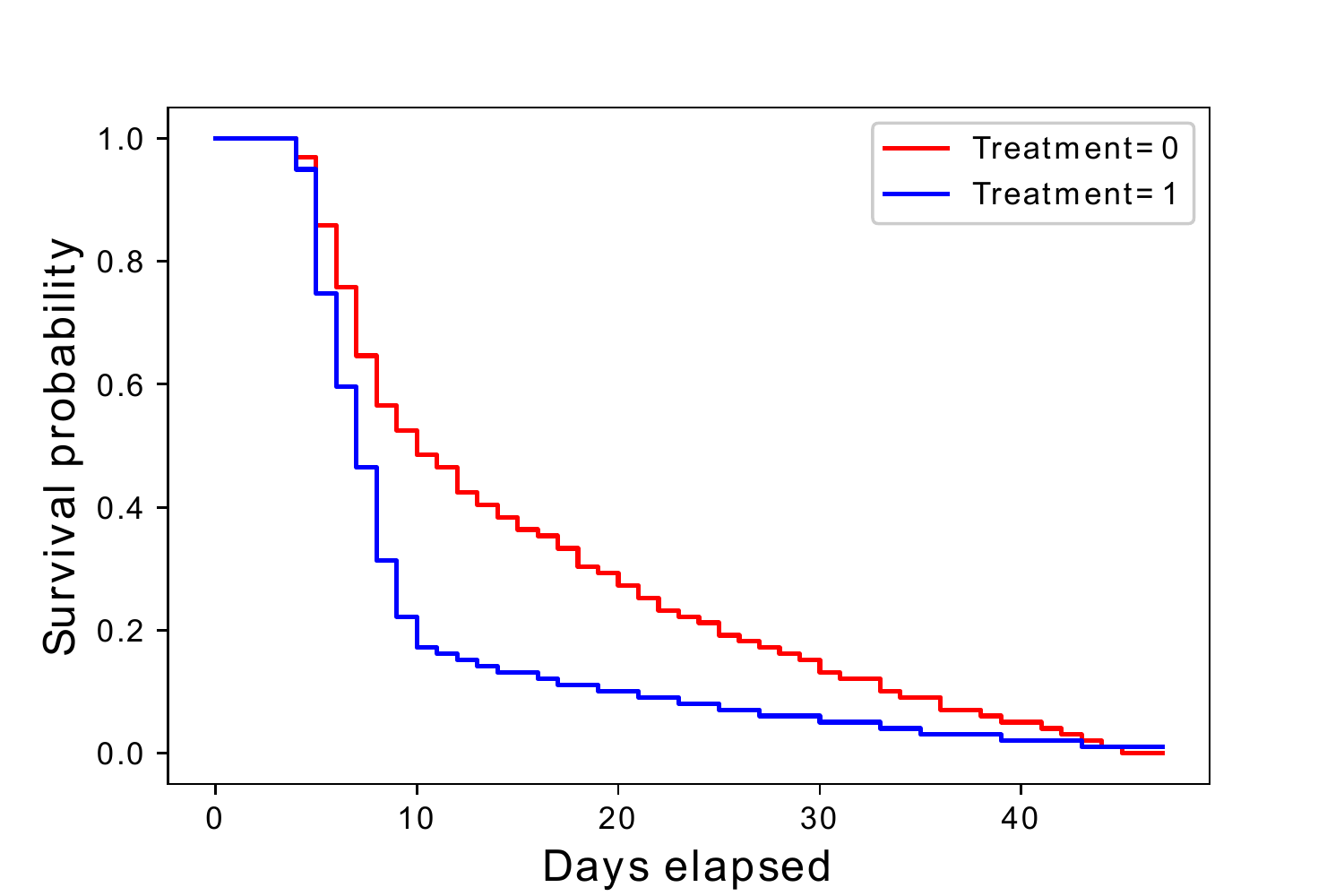}
    \end{minipage}
    \begin{minipage}{0.45\linewidth}
        \centering
        \includegraphics[width=\linewidth, trim = 10 1 30 34, clip]{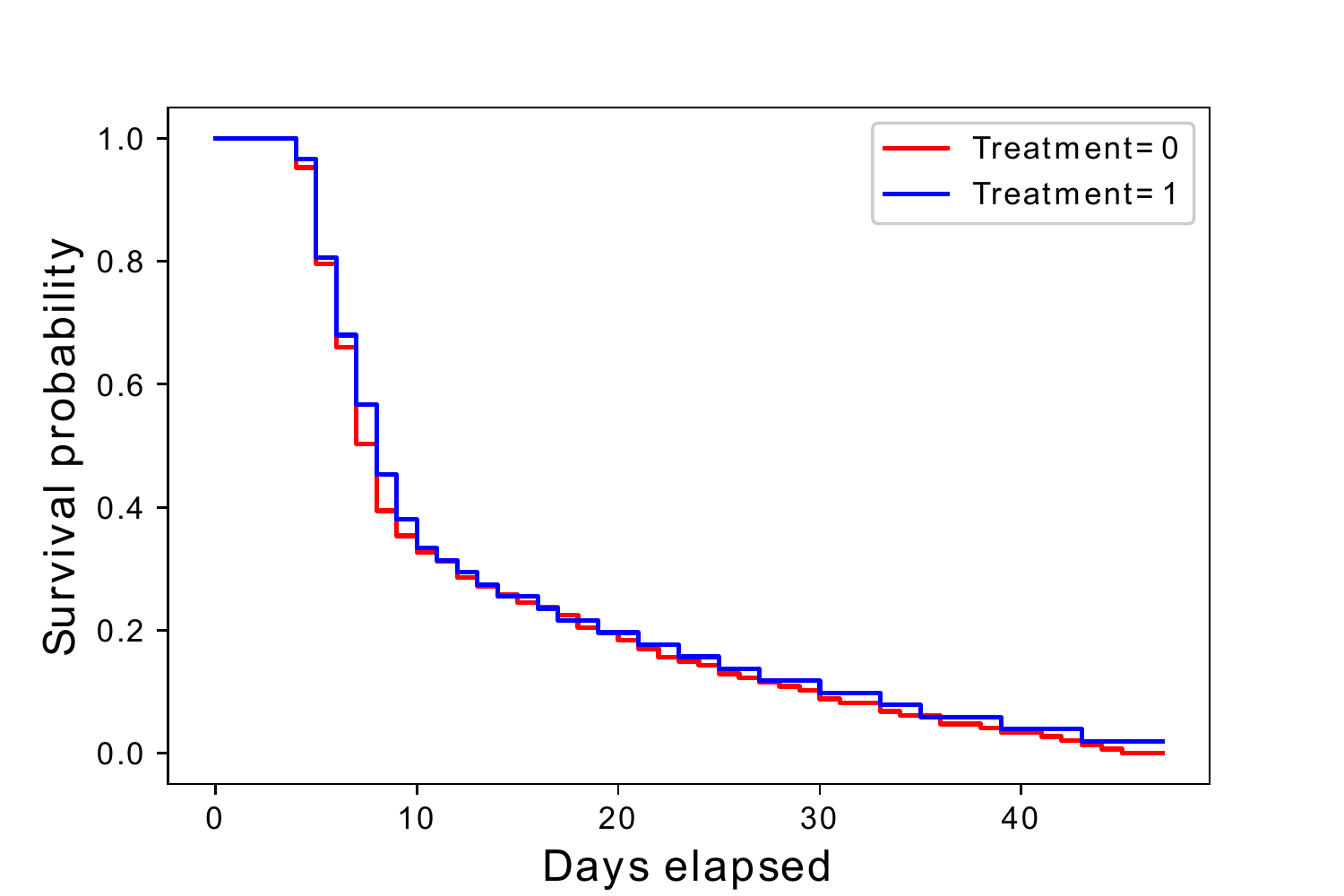}
    \end{minipage}
    \caption{Unadjusted survival curve for simulated data \textit{(left)} and, survival curve generated after applying proposed approach \textit{(right)}. Figure on left shows significant difference in survival curve between treatment and control group. The treatment population ($X=1$) seems to be more prone to hazard compared to the control population ($X=0$). Figure on right shows adjusted survival curve to be overlapping, signifying no significant difference in hazard rate in both the treatment and the control population.}
    \label{fig:surv_curve_sim}
\end{figure}

After we generate the experimental data, we applied Algorithm 1. We compared the existing approach of survival curve and survival curve from the adjusted dataset side-by-side in \autoref{fig:surv_curve_sim}.

\autoref{tab:table_hr_sim} presents the HR found in the fitted Cox PH model in three different processes:
\begin{enumerate}
    \item using only the treatment and outcome from the original dataset, 
    \item using data of all covariates (treatment, outcome and confounder) from the original dataset, and 
    \item using only the treatment and outcome from the adjusted dataset following our proposed approach. 
\end{enumerate}

The first approach represents scenarios where: 1) we ignore confounding, assuming it does not impact the treatment, or, 2) we do not possess data on the confounding variable (unmeasured confounding). This approach, however, results in an incorrect approximation of the HR. The second approach represents the existing approach to calculate HR. The third one shows our approach, and it eliminates the need for using confounding in model fitting since we are already adjusting for that.

\begin{table}[b!]
\centering
\resizebox{\textwidth}{!}{%
\begin{tabular}{lccc}
\hline
 & \multicolumn{2}{c}{Existing model} & Proposed model \\ \hline
 & \begin{tabular}[c]{@{}c@{}}Observational data \\ excluding confounding variable \\ (biased estimate)\end{tabular} & \begin{tabular}[c]{@{}c@{}}Observational data \\ including confounding variable \\ (traditional approach)\end{tabular} & \multicolumn{1}{l}{Transformed and adjusted data} \\ \hline
\multicolumn{1}{c}{Hazard Ratio} & \textbf{1.66} & \textbf{0.08} & \textbf{1.00} \\
\textit{(95\% Confidence Interval)} & \textit{\textbf{(1.25-2.20)}} & \textit{\textbf{(0.57-1.12)}} & \textit{\textbf{(0.76-1.33)}} \\ \hline
\end{tabular}%
}
    \caption{Hazard Ratio for simulated dataset, calculated using existing model and our proposed approach. 1st column reports a biased estimate of HR, by using only treatment and outcome (excluding confounder) in Cox PH model. The second column reports a standard estimate of HR, by including all known variables (including confounder). The third column reports HR calculated in our proposed approach, using only treatment and outcome (excluding confounder).}
    \label{tab:table_hr_sim}
\end{table}

Here, in \autoref{fig:surv_curve_sim}, the difference in unadjusted survival curve is similar to fitting Cox PH model with only $X$ and $T$, leaving out $Z$, as found following the first approach generating HR=1.66. On the other hand, the overlapping adjusted survival curve is validated by calculated HR, following both the existing approach with the Cox PH model (HR=0.8) and our algorithm (HR=1.0).
\begin{figure} [t]
    \centering
    \begin{minipage}{0.45\linewidth}
        \centering
        \includegraphics[width=\linewidth, trim = 10 1 30 34, clip]{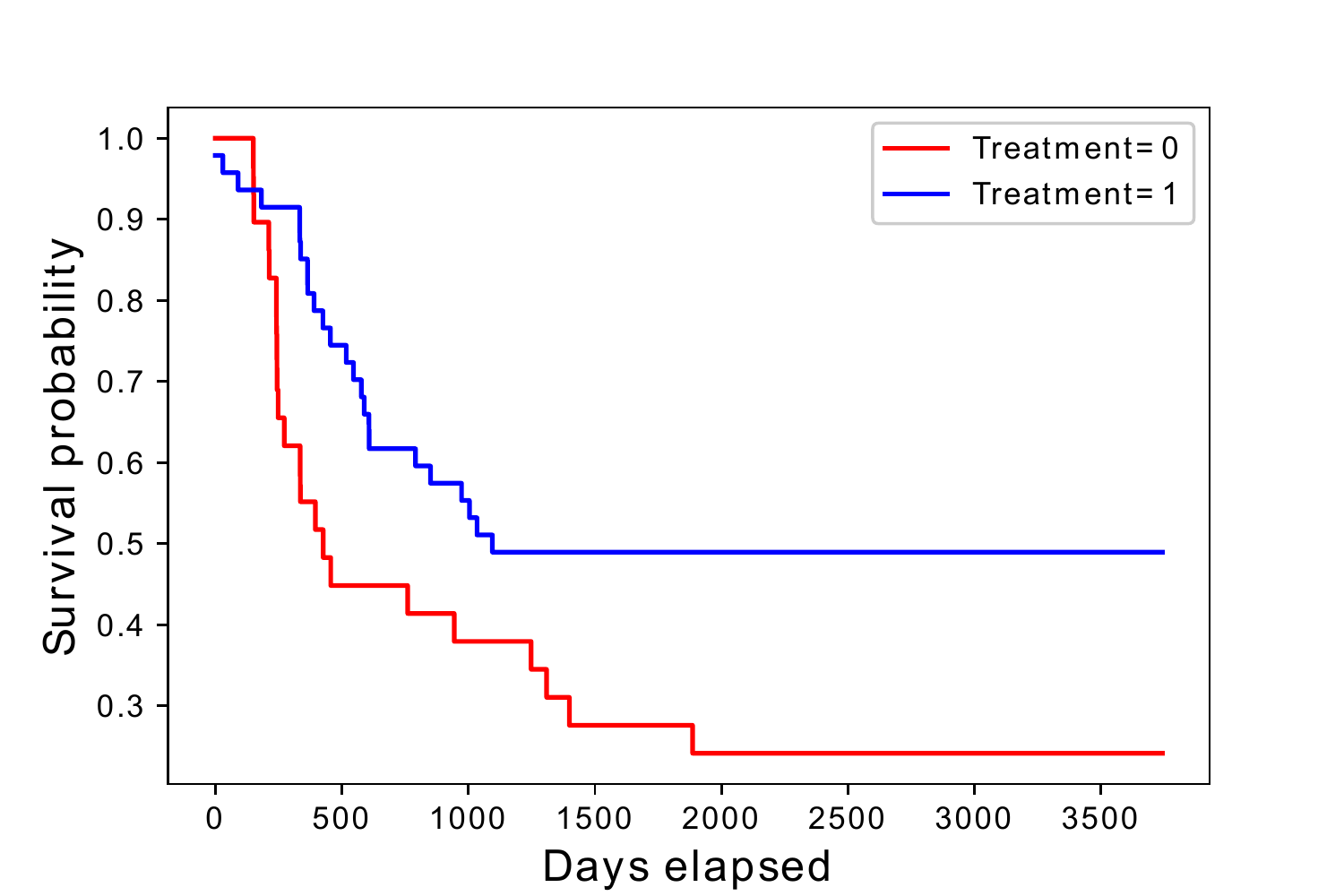}
    \end{minipage}
    \begin{minipage}{0.45\linewidth}
        \centering
        \includegraphics[width=\linewidth, trim = 10 1 30 34, clip]{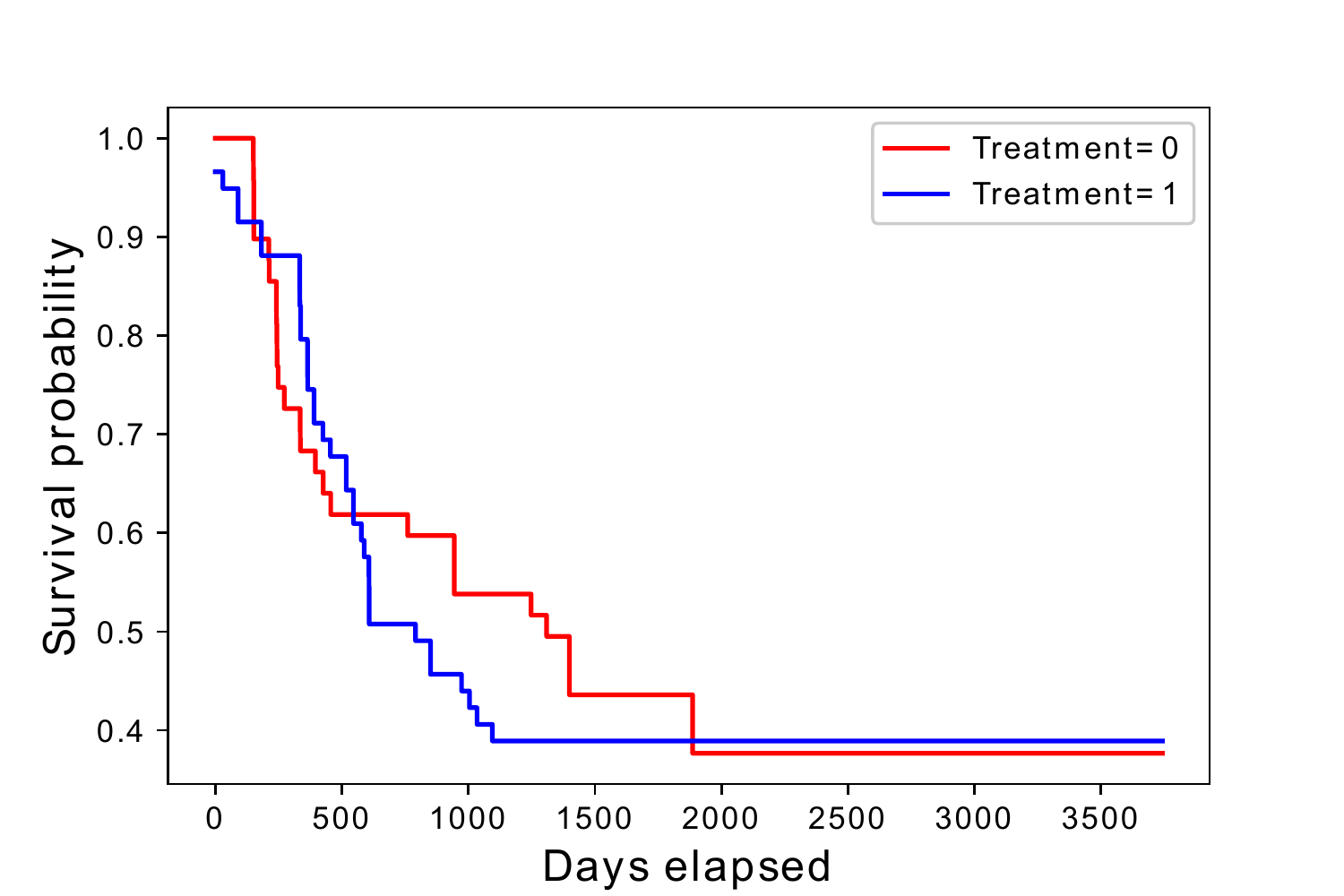}
    \end{minipage}
    \caption{Unadjusted survival curve for Ewing dataset \textit{(left)} and, survival curve generated after applying proposed approach \textit{(right)}. Figure on left presents treatment group ($X=1$) to be less hazardous than control group ($X=0$). Figure on right is the adjusted survival curve with mostly overlapping survival curves of two groups, although treatment group ($X=1$) seems slightly more prone to hazards.}
    \label{fig:surv_curve_ewing}
\end{figure}
\subsection{Ewing's Sarcoma Data}
We also applied the proposed method to a real-world dataset of patients with Ewing's Sarcoma \citep{makuch1982adjusted}. The dataset was selected based on its survival data and known causal DAG consisting of confounders.  The dataset consists of a total of 76 Ewing’s sarcoma patients with disease-free survival days as the outcome. 47 of the patients received a novel treatment (S4), and 29 received (one of) three (S1—S3) standard treatments. 

\begin{table}[b!]
\centering
\resizebox{\textwidth}{!}{%
\begin{tabular}{lccc}
\hline
 & \multicolumn{2}{c}{Existing model} & Proposed model \\ \hline
 & \begin{tabular}[c]{@{}c@{}}Observational data \\ excluding confounding variable \\ (biased estimate)\end{tabular} & \begin{tabular}[c]{@{}c@{}}Observational data \\ including confounding variable \\ (traditional approach)\end{tabular} & \multicolumn{1}{l}{Transformed and adjusted data} \\ \hline
\multicolumn{1}{c}{Hazard Ratio} & \textbf{0.53} & \textbf{1.12} & \textbf{1.04} \\
\textit{(95\% Confidence Interval)} & \textit{\textbf{(0.30-0.96)}} & \textit{\textbf{(0.59-2.11)}} & \textit{\textbf{(0.57-1.87)}} \\ \hline
\end{tabular}%
}
    \caption{Hazard Ratio for Ewing dataset, calculated using existing model and our proposed approach. The first column reports a biased estimate of the HR based on only the treatment and outcome (excluding confounder) in Cox PH model. The second column reports a standard estimate of the HR that includes all known variables (including the confounder(s)). The third column reports the HR calculated in our proposed approach, using only the treatment and outcome (excluding the confounder). The reason for getting an accurate estimate of the HR even when excluding the confounder is because we adjusted the dataset beforehand using a minimum set of confounders from the SCM, thus focusing on the true causal effect of treatment on outcome.}
    \label{tab:table_hr_ewing}
\end{table}

The level of Serum lactic acid dehydrogenase (LDH) acted as the confounder, since high LDH levels indicated a lesser likelihood of treatment assignment along with an impact on survival time. In our analysis, we marked patients receiving S4 as the treatment group ($X=1$) and patients receiving S1-S3 as the control group ($X=0$). We applied our algorithm on this data set and the survival curve with the existing approach. Results of our algorithm is shown in \autoref{fig:surv_curve_ewing}. The adjusted survival curve shows similar results, as demonstrated in Makuch et al. \citep{makuch1982adjusted}. We also present the calculated HR following the three processes described in the earlier subsection. In \autoref{tab:table_hr_ewing}, the HR calculated by our approach (HR=1.04) differs from the HR calculated in the traditional way (HR=1.12), presenting the drug to be a little less hazardous. However, the 95\% confidence interval for both of these coincide, signifying that the true value lies within this range.

\section{Discussion and Conclusion}

\begin{figure}[t]
    \centering
    \begin{minipage}{0.35\linewidth}
        \centering
        \includegraphics[width=\linewidth]{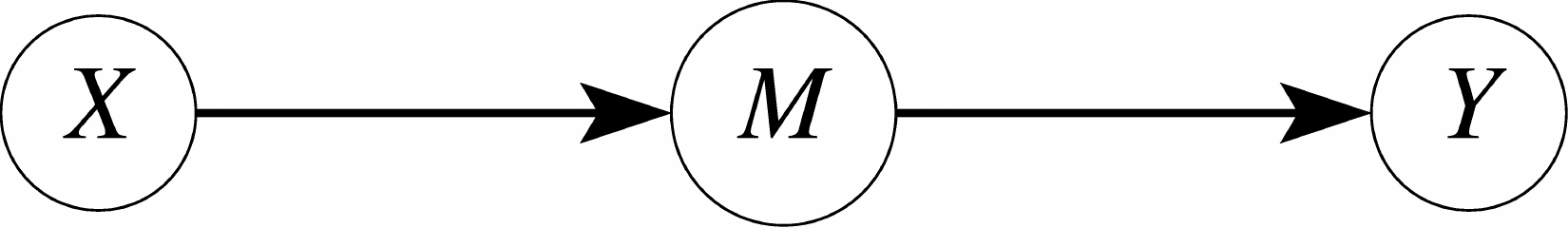}
    \end{minipage}
    \hspace{0.5cm}
    \begin{minipage}{0.35\linewidth}
        \centering
        \includegraphics[width=\linewidth]{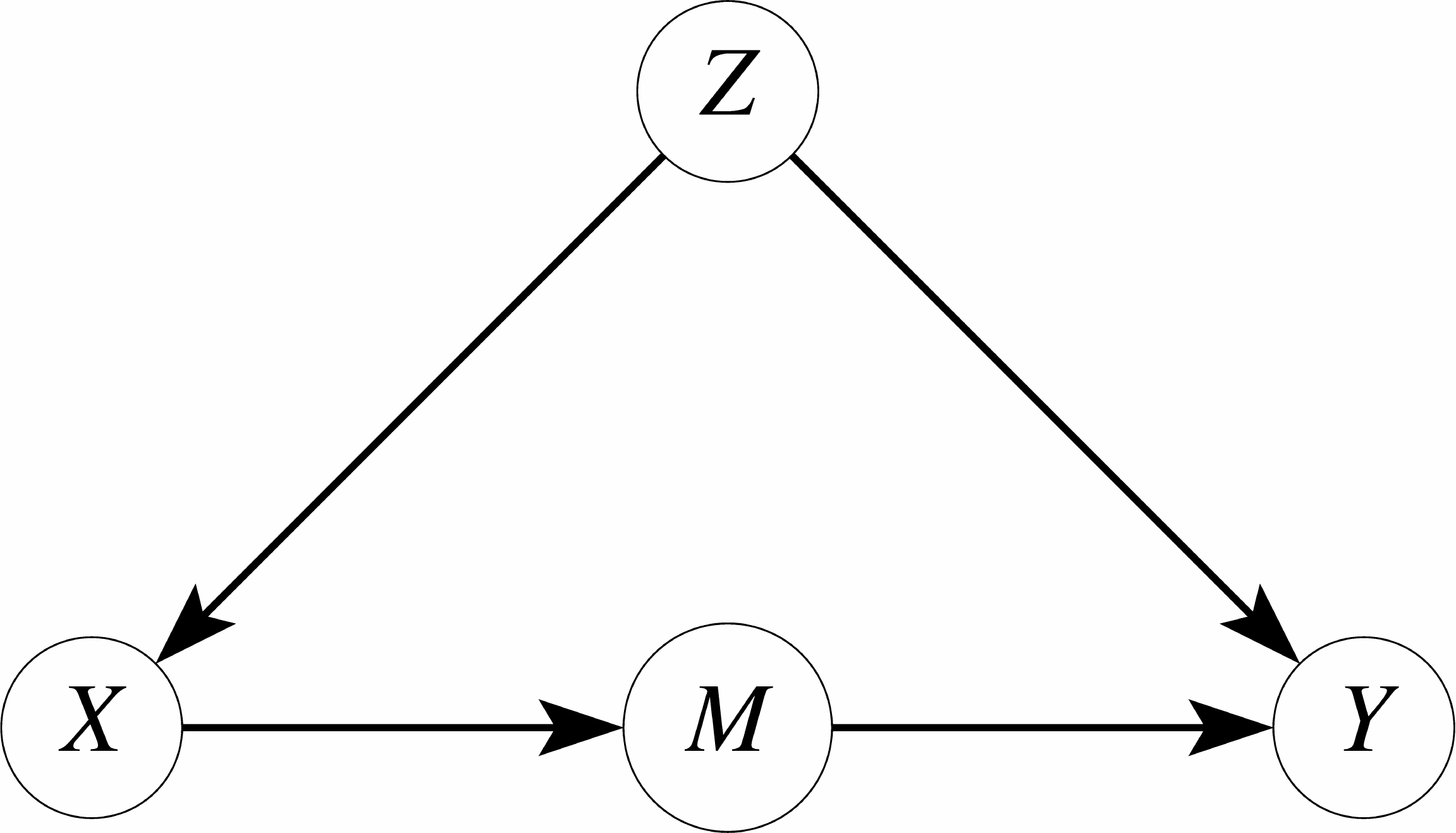}
    \end{minipage}
    \caption{Two example graphs where the backdoor adjustment will produce different results compared to an approach based on the ignorability assumption. In the left hand side, $X$ is the treatment, $Y$ is the mediator, and $M$ is a mediator. For the second graph, $Z$ acts as a confounder as well. The left hand side is the example of a mediator and the right hand side graph is known as the front-door setting.}
    \label{fig:causal_dag_2type}
\end{figure}

We propose a novel method to estimate the HR using the Cox PH Model through the transformation of observational data to corresponding experimental data leveraging an underlying SCM. The transformed data are mathematically guaranteed to be adjusted for the confounding bias with the assumption that the SCM represents the data generating mechanism. Previous approaches under RCM that estimate the survival curve use the ignorability assumption, and will not work when the variables selected do not satisfy the backdoor criterion. Ignorability assumption states that, distribution of the potential outcomes ($Y(0), Y(1)$) is independent of the treatment variable by randomly assigning treatment: $\{Y(0),Y(1)\} \indep X$. An extension of the idea, conditional ignorability states, distribution of the potential outcomes ($Y(0), Y(1)$) is independent of the treatment variable ($X$), conditional on the covariates ($Z$): $\{Y(0),Y(1)\} \indep X | Z$. Using conditional ignorability for adjustment on covariates allowed researchers to draw inferences from observational studies as well; however, adjusting all covariates irrespective of their causal relationship with treatment and outcome can contribute more bias to the model and incorrect estimation of causal effects.

We present two scenarios as example in \autoref{fig:causal_dag_2type}. In the first scenario (\autoref{fig:causal_dag_2type}, left), we show an SCM with treatment $X$ and outcome $Y$ with a third covariate $M$. Here $M$ acts as a mediator in between $X$ and $Y$, thus the backdoor adjustment gives a null set, meaning no adjustment is needed. The do-calculus formula would be: $P(Y|do(X)) = P(Y|X)$. Adjusting on $M$ based on conditional ignorability will produce an incorrect estimation of causal effect. In the second scenario (\autoref{fig:causal_dag_2type}, right), we discuss a setting called front-door adjustment where we identify the variables to be adjusted with two applications of backdoor adjustment \cite{pearl2009causality}. In an SCM with a mediator (shown in \autoref{fig:causal_dag_2type} (right)), the covariate $M$ does not satisfy the backdoor criterion and acts as a mediator between treatment $X$ and outcome $T$. Thus, adjusting with $M$ irrespective of its role as mediator will produce a biased estimate of the HR. The accurate backdoor adjustment formula (with $M$ as mediator) is $P(Y|do(X)) = \sum_M P(M|X) \sum_X P(Y|X,M) P(X)$. However, adjustment by  assuming $M$ (and $Z$) as confounder gives an incorrect adjustment formula: $P(Y|do(X)) = \sum_{M,Z} P(Y|M,X,Z) P(Z) P(M)$. Our approach (with backdoor criterion) can correctly identify the variables to be adjusted for estimating HR using SCM and do-calculus.  

Both the survival curve and the HR help to build a strong interpretation of the survival analysis of an experiment. The HR is most frequently reported as it summarizes the overall effect of treatment. However, survival curve encodes information on changes in survival over time \citep{hernan2010hazards}, which, in certain cases gives us better insight. Our proposed method is capable of generating both the survival curve and the HR, along with proper backdoor adjustment based on the underlying SCM. The HR calculated from the adjusted dataset requires only the treatment and outcome variables, and thus relies on direct causal relationships of treatment and outcome. For this purpose, we assume knowledge of the true causal model, an absence of unmeasured confounders, functional relationship in the SCM being time-invariant, and, proportionality of the HR in the outcome. In reality, defining the causal graph with SCM, that is, causal structure learning, requires a principled approach. The development of statistical and computational algorithms for causal structure learning is an active research area \citep{heinze2018causal,rottman2012causal}, and, is not well-established in the current literature. We are currently working on a methodological framework to develop the causal graph with structure learning algorithm and domain expertise. 

\section*{Acknowledgments}

This work is partially supported by a number of grants from the Regenstrief Center for Healthcare Engineering at Purdue University and Ubicomp Lab at Marquette University.


\bibliography{sample}



\end{document}